\documentclass[12pt]{article}
\usepackage{graphics,graphicx,graphpap,color}
\usepackage{setspace}

\newcommand\bmat{\left( \begin{array}{cc}}
\newcommand\emat{\end{array}\right)}

\def\msbar{\ifmmode{\overline{\rm MS}} \else{$\overline{\rm MS}$} \fi}
\def\drbar{\ifmmode{\overline{\rm DR}} \else{$\overline{\rm DR}$} \fi}

\def\ti              {\tilde}

\def\a               {\alpha}
\def\b               {\beta}
\def\d               {\delta}
\def\D               {\Delta}

\def\g               {\gamma}
\def\G               {\Gamma}

\def\t               {\theta}

\def\x               {\chi}

\def\sq              {{\ti q}}

\def\sqL             {{\ti q_L^{}}}
\def\sqR             {{\ti q_R^{}}}
\def\sf              {{\ti f}}

\def\st              {{\ti t}}
\def\sb              {{\ti b}}

\def\chp             {\ti \x^+}

\newcommand{\msq}[1]   {m_{\ti q_{#1}}}

\def\tw              {\t_{\scriptscriptstyle W}}

\def\non             {\nonumber}

\renewcommand\d{\delta}

\begin{document}
%---------------------------------------------------------------------

\pagestyle{empty} \vspace*{-1cm} 

\begin{flushright}
  HEPHY-PUB 771/03 \\
  hep-ph/0305250
\end{flushright}

\vspace*{2cm} 

\begin{center}
{\Large\bf\boldmath
   Improved full one--loop corrections \\[2mm] to {\boldmath{ $A^0 
   \rightarrow \tilde{q}_1\ {\bar{\!\tilde{q}}}_{2}$}} and 
   {\boldmath{ $\tilde{q}_2 \rightarrow \tilde{q}_1 A^0$}}}
   \\[5mm]

\vspace{10mm}

C.~Weber, H.~Eberl, W.~Majerotto\\[5mm]

\vspace{6mm}
{\it Institut f\"ur Hochenergiephysik der \"Osterreichischen 
Akademie der Wissenschaften, A--1050 Vienna, Austria} 

\end{center}

\vspace{20mm}

\begin{abstract}
We calculate the full electroweak one--loop corrections to the 
decay of the CP--odd Higgs boson $A^0$ into scalar quarks in the 
minimal supersymmetric extension of the Standard Model (MSSM). Due 
to the complex structure of the electroweak sector a proper 
renormalization of many parameters in the on--shell 
renormalization scheme is necessary. For the decay into sbottom 
quarks, especially for large $\tan\b$, the corrections can be very 
large in the on--shell renormalization scheme, which makes the 
perturbation series unreliable. We solve this problem by an 
appropriate definition of the tree--level coupling in terms of 
running quark masses and running trilinear couplings $A_q$. We 
also discuss the decay of heavy scalar quarks into light scalar 
quarks and $A^0$. We find that the corrections are significant and 
therefore cannot be neglected. 
\end{abstract}

\vfill
\newpage
\pagestyle{plain} \setcounter{page}{2}
%\baselineskip=17pt
%%%%%%%%%%%%%%%%%%%%%%%%%%%%%% Paper body %%%%%%%%%%%%%%%%%%%%%%%%

\section{Introduction}
The Minimal Supersymmetric Standard Model (MSSM) \cite{MSSM} 
requires five physical Higgs bosons: two neutral CP--even ($h^0$ 
and $H^0$), one heavy neutral CP--odd ($A^0$), and two charged 
ones ($H^\pm$) \cite{GunionHaber1, GunionHaber2}. The existence of 
a CP--odd neutral Higgs boson would provide a conclusive evidence 
of physics beyond the SM. Searching for Higgs bosons is one of the 
main goals of present and future collider experiments at TEVATRON, 
LHC or an $e^+ e^-$ Linear Collider. 

In this paper, we consider the decay of the CP--odd Higgs boson 
$A^0$ into two scalar quarks, $A^0 \rightarrow \tilde{q}_1 \ 
{\bar{\!\tilde{q}}}_{2}$. The decays into squarks can be the 
dominant decay modes of Higgs bosons in a large parameter region 
if the squarks are relatively light \cite{Bartl1, Bartl2}. In 
particular, the third generation squarks $\st_i$ and $\sb_i$ can 
be much lighter than the other squarks due to their large Yukawa 
couplings and their large left--right mixing. We will calculate 
the {\em full} electroweak corrections in the on--shell scheme and 
will implement the SUSY--QCD corrections which were calculated 
previously \cite{SUSY-QCD}. The challenge of this calculation is 
the necessity to renormalize almost all parameters in the 
electroweak sector in only one single process. Due to the numerous 
electroweak interacting particles and the complex coupling 
structure we have to compute a large number of graphs. In general, 
the Higgs--squark--squark couplings consist of $F$-- and 
$D$--terms and SUSY breaking terms, all depending on the squark 
mixing angle $\theta_\sq$. As a first step we consider the case 
$A^0 \rightarrow \tilde{q}_1 \ {\bar{\!\tilde{q}}}_{2}$ where only 
$F$--terms and SUSY breaking terms enter in the coupling. Since 
$A^0$ only couples to $\sq_L$--$\sq_R$ and due to the CP nature of 
$A^0$, $A^0 \rightarrow \tilde{q}_i \ {\bar{\!\tilde{q}}}_{i}$ 
vanishes (with real parameters also beyond the tree--level!). 
Despite the complexity, we have performed the calculation in an 
analytic way. The explicit formulae will be given elsewhere. We 
will, however, show the most important results of the numerical 
analysis. Furthermore, the crossed channel $\tilde{q}_2 
\rightarrow \tilde{q}_1 A^0$ is studied. 

In case of the decay into sbottom quarks the decay width can 
receive large corrections which makes the perturbation expansion 
unreliable, especially for large $\tan\b$. In some cases the width 
can even become negative using the on--shell renormalization 
scheme. We will show that this problem can be fixed by an 
appropriate choice of the tree--level coupling in terms of 
$\overline{\rm DR}$ running quark masses and running $A_q$. 

\vspace{2mm}
\section{Tree--level result}\label{treelevel}
The squark mixing is described by the squark mass matrix in the 
left--right basis $(\sqL, \sqR)$, and in the mass basis $(\sq_1, 
\sq_2)$, $\sq = \st$ or $\sb$, 
\begin{eqnarray}
  {\cal M}_{\sq}^{\,2} \,=\,
   \left( 
     \begin{array}{cc} 
       m_{\sq_L}^{\,2} & a_q\, m_q
       \\[2mm]
       a_q\,m_q & m_{\sq_R}^{\,2}
     \end{array}
   \right)
  = \left( R^\sq \right)^\dag
   \left( 
     \begin{array}{cc} 
       m_{\sq_1}^{\,2} & 0
       \\[2mm]
       0 & m_{\sq_2}^{\,2}
     \end{array}
   \right) R^\sq \,,
\end{eqnarray}
where $R^\sq_{i\a}$ is a 2 x 2 rotation matrix with rotation angle
$\theta_{\sq}$, which relates the mass eigenstates $\sq_i$, $i = 
1, 2$, $(m_{\sq_1} < m_{\sq_2})$ to the gauge eigenstates 
$\sq_\a$, $\a = L, R$, by $\sq_i = R^\sq_{i\a} \sq_\a$ and
\begin{eqnarray}
  m_{\sq_L}^{\,2} &=& M_{\ti Q}^2
       + (I^{3L}_q \!-\! e_{q}^{}\sin^2\!\tw)\cos2\b\,
       m_{\scriptscriptstyle Z}^{\,2}
       + m_{q}^2\,, \\[2mm]\label{MsD}
  m_{\sq_R}^{\,2} &=& M_{\{\ti U\!,\,\ti D \}}^2
       + e_{q}\sin^2\!\tw \cos2\b\,m_{\scriptscriptstyle Z}^{\,2}
       + m_q^2\,, \\[2mm]
  a_q &=& A_q - \mu \,(\tan\b)^{-2 I^{3L}_q} \,.
\end{eqnarray}
$M_{\ti Q}$, $M_{\ti U}$, and $M_{\ti D}$ are soft SUSY breaking 
masses, $A_q$ is the trilinear scalar coupling parameter, $\mu$ 
the higgsino mass parameter, $\tan\b = \frac{v_2}{v_1}$ is the 
ratio of the vacuum expectation values of the two neutral Higgs 
doublet states \cite{GunionHaber1, GunionHaber2}, $I^{3L}_q$ 
denotes the third component of the weak isospin of the quark $q$, 
$e_q$ the electric charge in terms of the elementary charge $e_0$, 
and $\tw$ is the Weinberg angle. 
\\ 
The mass eigenvalues and the mixing angle in terms of primary 
parameters are
\begin{eqnarray}
  \msq{1,2}^2  
    &=& \frac{1}{2} \left(
    \msq{L}^2 + \msq{R}^2 \mp
    \sqrt{(\msq{L}^2 \!-\! \msq{R}^2)^2 + 4 a_q^2 m_q^2}\,\right)
\\
  \cos\t_{\sq}
    &=& \frac{-a_q\,m_q}
    {\sqrt{(\msq{L}^2 \!-\! \msq{1}^2)^2 + a_q^2 m_q^2}}
  \hspace{2cm} (0\leq \t_{\sq} < \pi) \,,
\end{eqnarray}
and the trilinear breaking parameter $A_q$ can be written as
\begin{eqnarray}\label{mqAq}
m_q A_q  = \frac{1}{2} \left(m_{\sq_1}^2-m_{\sq_2}^2 \right) \sin 
2\theta_\sq \,+\, m_q \, \mu \,(\tan\b)^{-2 I^{3L}_q} \,. 
\end{eqnarray}
At tree--level the decay width of $A^0 \rightarrow \tilde{q}_1 \ 
{\bar{\!\tilde{q}}}_{2}$ is given by 
\begin{eqnarray}
\G^{\rm tree}(A^0 \rightarrow \tilde{q}_1 \ 
{\bar{\!\tilde{q}}}_{2}) &=& \frac{3\, \kappa (m_{A^0}^2, 
m^2_{\sq_1}, m^2_{\sq_2})}{16 \,\pi\, m^3_{A^0}}\ 
|G_{123}^{\sq}|^2 
\end{eqnarray}
with $\kappa (x, y, z) = \sqrt{(x-y-z)^2 - 4 y z}$ and the 
$A^0$--$\sq_i^\ast$--$\sq_j$ coupling $G_{ij3}^{\sq}$ given in 
\cite{SUSY-QCD}. 

\vspace{2mm}
\section{Full Electroweak Corrections}

\begin{figure}[th]
\begin{picture}(160,215)(0,0)
    %\graphpaper[5](0,0)(160,240)
    %\put(0,-1){\textcolor{red}{\framebox(159,215){}}}
     \put(0,-2){\mbox{\resizebox{16cm}{!}
     {\includegraphics{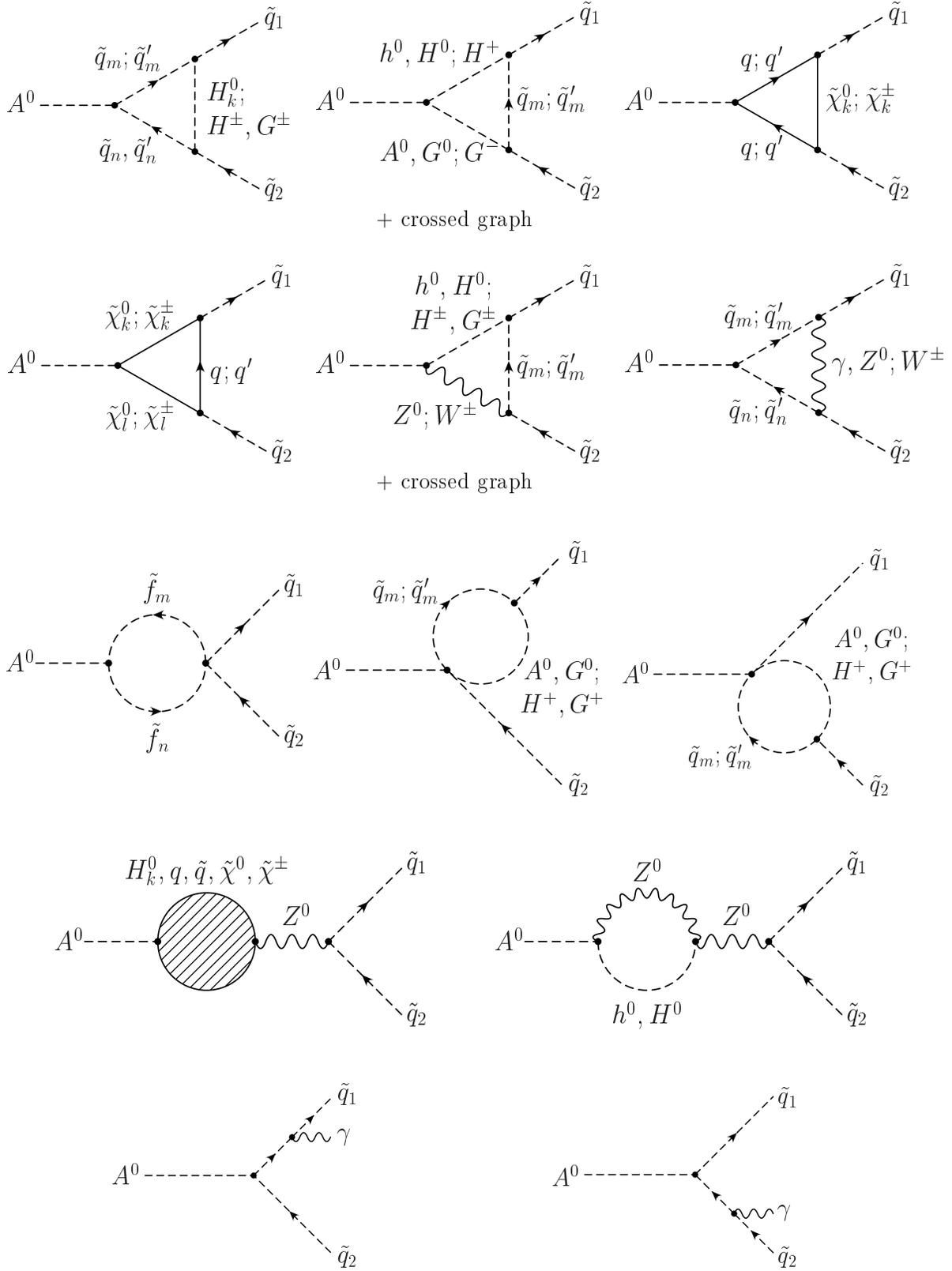}}}}
\end{picture}
\caption{Vertex and photon emission diagrams relevant to the 
calculation of the virtual electroweak corrections to the decay 
width $A^0 \rightarrow \tilde{q}_1 \ {\bar{\!\tilde{q}}}_{2}$. 
\label{vertex-graphs}} 
\end{figure}

\begin{figure}[th]
\begin{picture}(170,130)(0,0)
    %\graphpaper[5](0,0)(170,145)
     \put(0,0){\mbox{\resizebox{16cm}{!}
     {\includegraphics{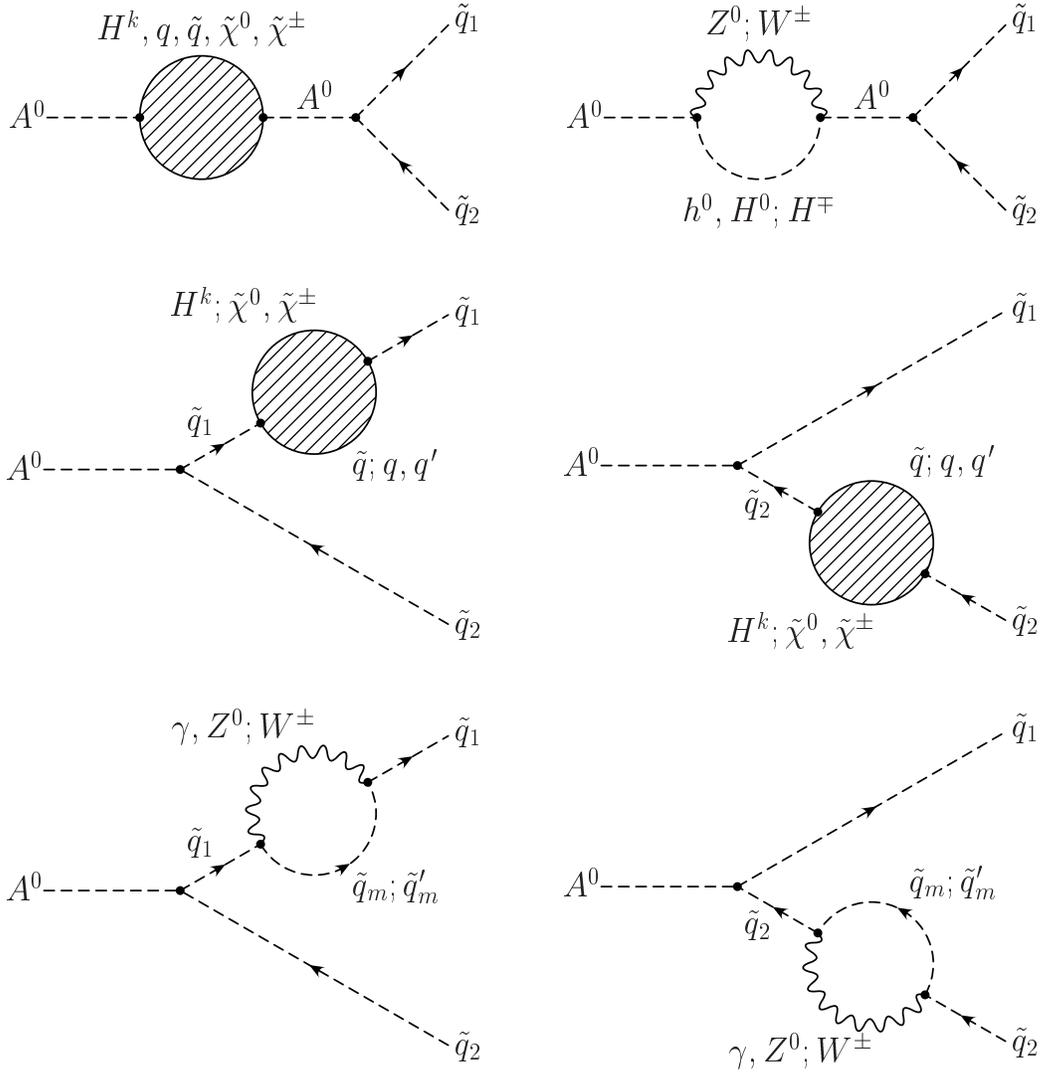}}}}
\end{picture}
\caption{Wave--function diagrams relevant to the calculation of 
the virtual electroweak corrections to the decay width $A^0 
\rightarrow \tilde{q}_1 \ {\bar{\!\tilde{q}}}_{2}$. $H^k$ denotes 
neutral as well as charged Higgs bosons. \label{WF-graphs}} 
\end{figure}

The one--loop corrected (renormalized) amplitude $G_{123}^{\sq\, 
\rm ren}$ can be expressed as 
\begin{eqnarray}
G_{123}^{\sq\, \rm ren} &=& G_{123}^{\sq} ~+~ \D G_{123}^{\sq} ~=~ 
G_{123}^{\sq} ~+~ \d G_{123}^{\sq (v)} ~+~ \d G_{123}^{\sq (w)} 
~+~ \d G_{123}^{\sq (c)} \,, 
\end{eqnarray}
where $\d G_{123}^{\sq (v)}$ are the vertex corrections
(Fig.~\ref{vertex-graphs}) and $\d G_{123}^{\sq (w)}$ the 
wave--function corrections (Fig.~\ref{WF-graphs}). Note that in 
addition to the one--particle irreducible vertex graphs also 
one--loop induced reducible graphs with $A^0$--$Z^0$ mixing have 
to be included. All parameters in the tree--level coupling 
$G_{123}^{\sq}$ have to be renormalized due to the shift from the 
bare to the on--shell values. These corrections are denoted by $\d 
G_{123}^{\sq (c)}$. The full one--loop corrected decay width is 
then given by 
\begin{eqnarray}\label{1loopwidth}
\G(A^0 \rightarrow \tilde{q}_1 \ {\bar{\!\tilde{q}}}_{2}) &=& 
\frac{3\, \kappa (m_{A^0}^2, m^2_{\sq_1}, m^2_{\sq_2})}{16 \,\pi\, 
m^3_{A^0}}\left[ |G_{123}^{\sq}|^2 + 2 {\rm Re} \left( 
G_{123}^{\sq}\cdot \D G_{123}^{\sq} \right) \right] \,. 
\end{eqnarray}
Since there are diagrams with photon exchange we also have to 
consider corrections due to real photon emission to cancel the 
infrared divergences (Fig.~\ref{vertex-graphs}). Therefore the 
corrected (UV-- and IR--convergent) decay width is 
\begin{eqnarray}\label{correctedwidth}
\G^{\rm corr}(A^0 \rightarrow \tilde{q}_1 \ 
{\bar{\!\tilde{q}}}_{2}) &\equiv& \G(A^0 \rightarrow \tilde{q}_1 \ 
{\bar{\!\tilde{q}}}_{2}) \,+\, \G(A^0 \rightarrow \tilde{q}_1 \ 
{\bar{\!\tilde{q}}}_{2}\,\g)\,. 
\end{eqnarray}
Throughout the paper we use the SUSY invariant dimensional 
reduction $(\overline{\rm DR})$ as regularization scheme. For 
convenience we perform the calculation in the 't Hooft--Feynman 
gauge, $\xi=1$.

\subsection{Vertex and wave--function corrections}
The relations between the unrenormalized (bare) and renormalized
(physical) fields and couplings are
\begin{eqnarray*}
      \qquad\mathcal L_0 &=& \mathcal L^{{\textrm{\scriptsize{ren}}}} +
      \delta\mathcal L \,,
\end{eqnarray*}

\begin{equation}\label{renormrelations}
   \begin{array}{l@{\qquad\qquad}l}
      \mathcal L_0 = \Big(G_{123}^\sq\Big)^0 \Big({A^0}\Big)^0
      \Big(\sq_1^{\ast}\Big)^0 \Big(\sq_2 \Big)^0 \,,
      & \Big( G_{123}^\sq \Big)^0 = G_{123}^\sq 
      + \delta G_{123}^{\sq(c)} \,,
      \\[3mm]
      \mathcal L^{{\textrm{\scriptsize{ren}}}} =
      G_{123}^\sq\,A^0\,\sq_1^\ast\sq_2 \,,
      & \Big({A^0}\Big)^0 = \sqrt{1 + \delta Z^H_{3k}}\ H_k^0 \,,
      \\[3mm]
      \delta\mathcal L = -\delta G_{123}^{\sq (v)} A^0 \sq_1^\ast \sq_2 \,,
      & \Big(\sq_1^{\ast}\Big)^0 = \sqrt{1 + \delta Z_{1i}^\sq}\
      \sq_{i}^{\ast} \,,
      \\[3mm]
      & \Big(\sq_2\Big)^0 = \sqrt{1 + \delta Z_{2j}^\sq}\ \sq_{j} \,,
   \end{array}
\end{equation}
with the notation $H_k^0 = \{h^0, H^0, A^0, G^0\}$, $i, j = 1, 2$, 
and $k = 3, 4$. Thus the renormalized Lagrangian is given by (up 
to the first order) 
\begin{eqnarray}\non
   \mathcal L^{{\textrm{\scriptsize{ren}}}} & = & \left(G_{123}^\sq +
   \delta G_{123}^{\sq (v)} + \frac{1}{2}\Big(\delta Z_{i1}^\sq G_{i23}^\sq +
   \delta Z_{j2}^\sq G_{1j3}^\sq + \delta Z_{k3}^H G_{12k}^\sq \Big)
   + \delta G_{123}^{\sq (c)} \right)A^0 \sq_1^\ast \sq_2 \,.
   \\
\end{eqnarray}
The explicit form of the vertex corrections $\delta G_{123}^{\sq 
(v)}$ will be given elsewhere. Due to the antisymmetry of the 
tree--level coupling, $G^\sq_{ij3} = -G^\sq_{ji3}$, the total
wave--function correction reads
\begin{eqnarray}
   \delta G_{123}^{\sq (w)} &=& \frac{1}{2}\Big(\delta Z_{11}^\sq 
   + \delta Z_{22}^\sq + \delta Z_{33}^H \Big) G_{123}^\sq
   + \frac{1}{2}\delta Z_{43}^H G_{124}^\sq \,.
\end{eqnarray}
For the wave--function renormalization constants we use the 
conventional on--shell renormalization conditions 
\cite{onshellren} which lead to 
\begin{equation}
   \begin{array}{l}
      \delta Z_{ii}^{\sq} ~=~ - \Re\,\dot\Pi_{ii}^{\sq} (m_{\sq_i}^2)\ ,
      \\[3mm]
      \delta Z_{33}^{H} ~=~ - \Re\,\dot\Pi_{33}^{H} (m_{A^0}^2)\ ,
   \end{array}\qquad
   \delta Z_{43}^{H} ~=~ \frac{2}{m_{G^0}^2\!-\!m_{A^0}^2}\, \Re\,\Pi_{43}^{H}
   (m_{A^0}^2) \,, \qquad
\end{equation}
with the diagonal parts of the Higgs and squark self--energies
$\dot\Pi_{ii}(k^2)$.

%%%%%%%%%%%%%%%%%%%%%%%%%%%%%%% Fig 1 %%%%%%%%%%%%%%%%%%%%%%%%%%%%%%%%%
%%%%%%%%%%%%%%%%%%%%%%%%%%%%%%%%%%%%%%%%%%%%%%%%%%%%%%%%%%%%%%%%%%%%%%%
\clearpage
 
%%%%%%%%%%%%%%%%%%%%%%%%%%%%%%% Fig 2 %%%%%%%%%%%%%%%%%%%%%%%%%%%%%%%%%
%%%%%%%%%%%%%%%%%%%%%%%%%%%%%%%%%%%%%%%%%%%%%%%%%%%%%%%%%%%%%%%%%%%%%%%

\noindent The off--diagonal Higgs wave--function corrections can 
be combined with the contribution to $\d G_{123}^{\sq (v)}$ which 
come from $A^0$--$Z^0$ mixing. First we show that the sum of the 
parts coming from the propagators of $Z^0$ and $G^0$ outside the 
loops is independent of the gauge \mbox{parameter $\xi = \xi_Z$.} 

%%%%%%%%%%%%%%%%%%%%%%%%%%%%%%% Fig 3 %%%%%%%%%%%%%%%%%%%%%%%%%%%%%%%%%
\begin{figure}[h!]
\begin{picture}(170,35)(0,0)
    %\graphpaper[5](0,0)(170,35)
     \put(0,0){\mbox{\resizebox{16cm}{!}
     {\includegraphics{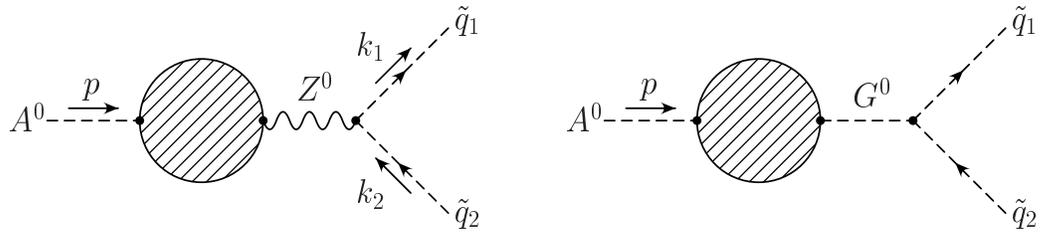}}}}
\end{picture}
\caption{$A^0$--$Z^0$ contribution and $A^0$--$G^0$ wave--function 
correction\label{GI}} 
\end{figure}
%%%%%%%%%%%%%%%%%%%%%%%%%%%%%%%%%%%%%%%%%%%%%%%%%%%%%%%%%%%%%%%%%%%%%%%

\noindent The amplitudes of the two graphs of Fig.~\ref{GI} in a 
general $R_\xi$ gauge are 
\begin{eqnarray}
   \mathcal M^Z &=& \Big(\! -i p^\mu \Pi_{AZ}(p^2) \Big) 
   \frac{i}{p^2\!-\!m_{\scriptscriptstyle Z}^2}
   \bigg(\! -\!g_{\mu\nu}+(1\!-\!\xi)\frac{p_\mu p_\nu}{p^2\!-\!\xi 
   m_{\scriptscriptstyle Z}^2} \bigg)
   \Big(\! -i g_Z\, z^\sq_{12} \Big) (k_1\!+\!k_2)^\nu\,, \hspace{10mm}
   \\[3mm]
   \mathcal M^G &=& \Big( i \Pi_{AG}(p^2) \Big) \frac{i}{p^2\!-\!\xi 
   m_{\scriptscriptstyle Z}^2}\,
   i G^\sq_{124}\,.
\end{eqnarray}
Contracting the Lorentz indices in $\mathcal M^Z$,
\begin{eqnarray}
   p^\mu \bigg(\! -\!g_{\mu\nu}+(1\!-\!\xi)\frac{p_\mu p_\nu}{p^2\!-\!\xi 
   m_{\scriptscriptstyle Z}^2}
   \bigg) (k_1\!+\!k_2)^\nu &=& - \bigg( 1-\frac{(1-\xi)p^2}{p^2 - \xi 
   m_{\scriptscriptstyle Z}^2} 
   \bigg) \Big( m_{\sq_1}^2 - m_{\sq_2}^2 \Big)\,,
\end{eqnarray}
and eliminating $\Pi_{AG}$ in favor of $\Pi_{AZ}$ by using the 
Slavnov--Taylor identity \cite{dabelstein} 
\begin{eqnarray}
   p^2 \Pi_{AZ} (p^2) + i m_{\scriptscriptstyle Z} \Pi_{AG}(p^2) = 0\,,
\end{eqnarray}
we find the sum $\mathcal M^Z + \mathcal M^G$
\begin{eqnarray}\non
   \mathcal M^{Z+G} &=& \frac{i}{p^2\!-\!m_{\scriptscriptstyle Z}^2} 
   \Pi_{AZ}(p^2)\, g_Z\, z^\sq_{12}
   \Big( m_{\sq_1}^2 - m_{\sq_2}^2 \Big) \bigg( 1-\frac{(1-\xi)p^2}{p^2 - 
   \xi m_{\scriptscriptstyle Z}^2} \bigg) 
   \\[2mm]
   && +~\frac{p^2}{p^2 - \xi m_{\scriptscriptstyle Z}^2}\, 
   \frac{\Pi_{AZ}(p^2)}{m_{\scriptscriptstyle Z}}\, G^\sq_{124}\,.
\end{eqnarray}
Finally we use the identity
\begin{eqnarray}
   g_Z \, z^\sq_{ij}\, \Big( m_{\sq_i}^2 - m_{\sq_j}^2 \Big) 
   &=& i m_{\scriptscriptstyle Z} \,G^\sq_{ij4}
\end{eqnarray}
to obtain the result 
\begin{eqnarray}\non
   \d G_{123}^{\sq (Z+G)} &=& -i \mathcal M^{Z+G} 
   (p^2\rightarrow m_{A^0}^2)
   ~=~ -\frac{i \, \Pi_{AZ}(m_{A^0}^2)\, 
   G^\sq_{124}}{m_{\scriptscriptstyle Z}\, 
   (p^2\!-\!m_{\scriptscriptstyle Z}^2)\,
   (p^2\!-\!\xi m_{\scriptscriptstyle Z}^2)} \,\times
   \\[3mm]\non
   &&\bigg[ -m_{\scriptscriptstyle Z}^2 \bigg( 
   \Big( p^2 - \xi m_{\scriptscriptstyle Z}^2 \Big)
   -(1-\xi) p^2 \bigg) + p^2 \Big( p^2 - m_{\scriptscriptstyle Z}^2 
   \Big)\bigg]
   \\[3mm]
   &=& - \frac{i}{m_{\scriptscriptstyle Z}}\, 
   \Pi_{AZ}(m_{A^0}^2)\, G^\sq_{124} \,.
\end{eqnarray}
The gauge dependence of the propagators of the $Z^0$ and $G^0$ in 
Fig.~\ref{GI} is completely removed. However, there still remain
gauge dependences from vector particles and Goldstone bosons in 
the loops of $\Pi_{AZ}$ which cancel against their counter parts 
in the vertex, wave--function and counter term corrections.

%\clearpage
\subsection{Counter terms}
Since all parameters in the tree--level coupling $G^\sq_{123}$ 
have to be renormalized, we get 
\begin{eqnarray} \d G_{123}^{\sq (c)} &=& 
\frac{\d h_q}{h_q}\, G_{123}^{\sq} + \frac{i}{\sqrt2}\, h_q \ \d 
\! \left( \hspace{-2pt} {A_q} \left\{ \hspace{-5pt} 
\begin{array}{rr}{\cos\beta} \\ {\sin\beta}
\end{array} \hspace{-5pt} \right\} + \mu \left\{ \hspace{-5pt}
\begin{array}{rr}{\sin\beta} \\
{\cos\beta} \end{array} \hspace{-5pt} \right\} \right)
\end{eqnarray}
for 
{\scriptsize{$\left\{\hspace{-4pt}\begin{array}{cc}{\textrm{up}}\\ 
{\textrm{down}}\end{array}\hspace{-4pt}\right\}$}}--type squarks.
The Yukawa coupling counter term can be decomposed into 
corrections to the electroweak coupling $g$, the masses of the 
quark $q$ and the gauge boson $W$ and the mixing angle $\b$, 
\begin{eqnarray}
   \frac{\d h_q}{h_q} &=& \frac{\d g}{g} \,+\, \frac{\d m_q}{m_q} 
   \,-\, \frac{\d m_{\scriptscriptstyle W}}{m_{\scriptscriptstyle W}} 
   \,+\, {\left\{ \hspace{-5pt}
   \begin{array}{r}{-\cos^2\beta} \\ {\sin^2\beta}
   \end{array} \hspace{-3pt} \right\}}
   \frac{\d\tan\beta}{\tan\beta} \,.
\end{eqnarray}
For the trilinear coupling we get with eq.~(\ref{mqAq})
\begin{eqnarray}\label{dAfrel}
   \frac{\d A_q}{A_q} &=& \frac{\d (m_q A_q)}{m_q A_q} -
   \frac{\d m_q}{m_q} \,,
\\[2mm] \non
   \d(m_q A_q) &=& \d \!\left(\! m_q \mu\, {\left\{ \hspace{-5pt}
   \begin{array}{rr}{\cot\beta} \\ {\tan\beta}
   \end{array} \hspace{-5pt} \right\}} \!\right) \,+\, \frac{1}{2}\!
   \left( \d m_{\sq_1}^2 \!-\! \d m_{\sq_2}^2 \right)\sin 2\theta_\sq
\\[2mm]\label{dmqAq}
   &&+ \left(  m_{\sq_1}^2 \!-\! m_{\sq_2}^2 \right)
   \cos 2\theta_\sq \, \d\theta_\sq \,.
\end{eqnarray}
In the on--shell scheme the renormalization condition for the 
electroweak gauge boson sector reads \cite{sirlin}
\begin{eqnarray} 
   \frac{\d g}{g} &=& \frac{\d e}{e} \,+\, \frac{1}{\tan^2\tw}
   \left( \frac{\d m_{\scriptscriptstyle W}}{m_{\scriptscriptstyle W}}
   - \frac{\d m_{\scriptscriptstyle Z}}{m_{\scriptscriptstyle Z}} \right)
\end{eqnarray}
with $m_{\scriptscriptstyle W}$ and $m_{\scriptscriptstyle Z}$ 
fixed as well as the quark and squark masses as the physical 
(pole) masses.\\ 

\noindent{\bf Renormalization of the electric charge 
{\boldmath{$e$}}}\\ Since we use as input parameter for $\a$ the 
$\overline{\rm MS}$ value at the $Z$--pole, $\a \equiv 
\a(m_{\scriptscriptstyle Z})|_{\overline{\rm MS}} = e^2/(4\pi)$, 
we get the counter term \cite{0111303} 
\begin{eqnarray}\non
  \frac{\d e}{e} &=& \frac{1}{(4\pi)^2}\,\frac{e^2}{6} \Bigg[
  \,4 \sum_f N_C^f\, e_f^2 \bigg(\D + \log\frac{Q^2}{x_f^2} \bigg)
  + \sum_{\sf} \sum_{m=1}^2 N_C^f\, e_f^2
  \bigg( \D + \log\frac{Q^2}{m_{\sf_m}^2} \bigg)
  \\ \non
  && \hspace{18mm}
  -4 \sum_{k=1}^2 \bigg( \D + \log\frac{Q^2}{m_{\chp_k}^2} \bigg)
  - \sum_{k=1}^2 \bigg( \D + \log\frac{Q^2}{m_{H_k^+}^2} \bigg)
  - 2 \bigg( \D + \log\frac{Q^2}{m_{\scriptscriptstyle W}^2} \bigg) 
  \Bigg] \,.
  \\
\end{eqnarray}
with $x_f = m_{\scriptscriptstyle Z} \ \forall\ m_f < 
m_{\scriptscriptstyle Z}$ and $x_t = m_t$.  $N_C^f$ is the colour 
factor, $N_C^f = 1, 3$ for (s)leptons and (s)quarks, respectively. 
$\D$ denotes the UV divergence factor, \mbox{$\D = 2/\epsilon - \g 
+ \log 4\pi$}. 
\\ 

\noindent{\bf Renormalization of {\boldmath{$\tan\b$}}}\\ For 
$\tan\b$ we use the condition \cite{pokorski} Im$\hat\Pi_{A^0 
Z^0}(m_A^2) = 0$  which gives the counter term 
\begin{eqnarray}\label{dtanb}
   \frac{\d \tan\b}{\tan\b} &=& \frac{1}{m_{\scriptscriptstyle Z} 
   \sin 2\b}\, {\rm Im} 
   \Pi_{A^0 Z^0} (m_{A^0}^2).
\end{eqnarray}\\

\noindent{\bf Renormalization of {\boldmath{$\mu$}}}\\ The 
higgsino mass parameter $\mu$ is renormalized in the chargino 
sector \cite{0104109, Willi} where it enters in the 22--element of 
the chargino mass matrix $X$, 
\begin{eqnarray}\label{renmu}
   X = \left(
   \begin{array}{cc}
      M & \sqrt2 m_{\scriptscriptstyle W} \sin\b \\ \sqrt2 
  m_{\scriptscriptstyle W} \cos\b & {\mu}
   \end{array}\right) \hspace{10mm}
   \rightarrow {\d\mu ~=~ (\d X)_{22}} \,.
\end{eqnarray}\\

\noindent{\bf Renormalization of {\boldmath{$\theta_{\!\sq}$}}}\\ 
The counter term of the squark mixing angle, $\d\theta_{\!\sq}$, 
is fixed such that it cancels the anti--hermitian part of the 
squark wave--function corrections \cite{guasch, JHEP9905}, 
\begin{eqnarray}\label{dthetasf}
   \delta \theta_{\sq} & = & \frac{1}{4}\, \left(
   \d Z^\sq_{12} - \d Z^\sq_{21}\right)
   = \frac{1}{2\big(m_{\sq_1}^2 \!-\! m_{\sq_{2}}^2\big)}\, {\rm Re}
   \left( \Pi_{12}^\sq(m_{\sq_{2}}^2) + \Pi_{21}^\sq(m_{\sq_{1}}^2)
    \right) \,.
\end{eqnarray}

\subsection{Infrared divergences}
The infrared divergences in eq.~(\ref{1loopwidth}) are cancelled 
by the inclusion of real photon emission, see the last two Feynman 
diagrams of Fig.~\ref{vertex-graphs}. The decay width of $A^0(p) 
\rightarrow \tilde{q}_1(k_1) +\ {\bar{\!\tilde{q}}}_{2}(k_2) + 
\g(k_3)$ can be written as 
\begin{eqnarray}\non
\G(A^0 \rightarrow \tilde{q}_1 \, {\bar{\!\tilde{q}}}_{2}\,\g) &=& 
\frac{3(e\,e_q)^2\,|G_{123}^{\sq}|^2}{16\,\pi^3\,m_{A^0}} \left[ 
\left( m_{A^0}^2\!-\!m_{\sq_1}^2\!-\!m_{\sq_2}^2 \right) I_{12} 
\!-\!m_{\sq_1}^2 I_{11}\!-\!m_{\sq_2}^2 
I_{22}\!-\!I_1\!-\!I_2\right] 
\\  
\end{eqnarray}
with the phase--space integrals $I_n$ and $I_{mn}$ defined as 
\cite{Denner} 
\begin{eqnarray}
I_{i_1\ldots i_n}=\frac{1}{\pi^2}
\int\frac{d^3k_1}{2E_1}\frac{d^3k_2}{2E_2}\frac{d^3k_3}{2E_3}
\delta^4(p-k_1-k_2-k_3)\frac{1}
{(2k_3k_{i_1}+\lambda^2)\ldots(2k_3k_{i_n}+\lambda^2)}.
\end{eqnarray}
The corrected (UV-- and IR--convergent) decay width is then given 
by (see eq.~(\ref{correctedwidth})) 
\begin{eqnarray}
\G^{\rm corr}(A^0 \rightarrow \tilde{q}_1 \,
{\bar{\!\tilde{q}}}_{2}) &\equiv& \G(A^0 \rightarrow \tilde{q}_1 
\, {\bar{\!\tilde{q}}}_{2}) \,+\, \G(A^0 \rightarrow \tilde{q}_1 
\, {\bar{\!\tilde{q}}}_{2}\,\g)\,. 
\end{eqnarray}

\section{Improvement of One--loop Corrections}
In the on--shell renormalization scheme, in case of the decay into 
sbottom quarks, especially for large $\tan\b$, the decay width can 
receive large corrections which makes the perturbation expansion 
unreliable. In some cases the corrected width can even become 
negative. It has been pointed out \cite{dmb, impSUSYQCD} that the 
source of these large corrections are mainly the counter terms for 
$m_b$ and the trilinear coupling $A_b$. We show that this problem 
can be fixed by absorbing these large counter terms into the 
$A^0$--squark--squark tree--level coupling and expanding the 
perturbation series around the new tree--level. The technical 
details will be given in a forthcoming paper. 
\\ 

\noindent{\bf Correction to 
{\boldmath{$m_b$}\label{sec:mbMSSM}}}\\ If the Yukawa coupling 
$h_b$ is given at tree--level in terms of the pole mass $m_b$, the 
one--loop corrections to the counter term $\d m_b$ become very 
large due to gluon and gluino exchange contributions. We absorb 
these large counter terms and also the ones due to loops with 
electroweak interacting particles into the Higgs--squark--squark 
tree--level coupling by using the $\overline{\rm DR}$ running  
mass $\hat m_b (Q\!=\!m_A)$. The large counter term due to the 
gluon loop is absorbed by using SM 2--loop renormalization group 
equations \cite{impSUSYQCD, BL, hdecay}. Thus we obtain the SM 
running bottom $\hat m_b(Q)_{\rm SM}$. For large $\tan\b$ the 
counter term to $m_b$ can be very large due to the 
gluino--mediated graph \cite{dmb, hbbnew, chankowski}. Here we 
absorb the gluino contribution as well as the sizeable 
contributions from neutralino and chargino loops and the remaining 
electroweak self--energies into the Higgs--squark--squark 
tree--level coupling. In such a way we obtain the full 
$\overline{\rm DR}$ running bottom quark mass 
\begin{eqnarray}\label{mbMSSM}
   \hat m_b(Q)_{\rm MSSM} &=& \hat m_b(Q)_{\rm SM} + \d m_b(Q)\,.
\end{eqnarray}

\noindent{\bf Correction to {\boldmath{$A_b$}}}\\ The second 
source of a very large correction (in the on--shell scheme) is the 
counter term for the trilinear coupling $A_b$, eqs.~(\ref{dAfrel}, 
\ref{dmqAq}), especially the contribution of the left--right 
mixing elements of the squark mass matrix, $m_{\scriptscriptstyle 
LR}^2 = (m_{\sq_1}^2 - m_{\sq_2}^2) \sin \theta_\sq \cos 
\theta_\sq$. As in the case of the large correction to $m_b$ we 
use $\overline{\rm DR}$ running $\hat A_b (m_{A^0})$ in the 
Higgs--squark--squark tree--level coupling. Because of the fact 
that the counter term $\d A_b$ (for large $\tan\b$) can become 
several orders of magnitude larger than the on--shell $A_b$ we use 
$\hat A_b (m_{A^0})$ as input \cite{impSUSYQCD}. In order to be 
consistent we have to perform an iteration procedure to get the 
correct running and on--shell masses, mixing angles and other 
parameters.

\section{Numerical analysis and conclusions}
In the following numerical examples, we assume $M_{{\ti Q}} \equiv 
M_{{\ti Q}_{3}} = \frac{10}{9} M_{{\ti U}_{3}} = \frac{10}{11} 
M_{{\ti D}_{3}} = M_{{\ti L}_{3}} = M_{{\ti E}_{3}} = M_{{\ti 
Q}_{1,2}} = M_{{\ti U}_{1,2}} = M_{{\ti D}_{1,2}} = M_{{\ti 
L}_{1,2}} = M_{{\ti E}_{1,2}}$ for the first, second and third 
generation soft SUSY breaking masses and $A \equiv A_t = A_b = 
A_\tau$, if not stated otherwise. For the standard model 
parameters we take $m_{\scriptscriptstyle Z} = 91.1876$~GeV, 
$m_{\scriptscriptstyle W} = 80.423$~GeV, $\sin^2 \tw = 1 - 
m_{\scriptscriptstyle W}^2/m_{\scriptscriptstyle Z}^2$, $\a = 
1/127.934$, $m_t = 174.3$~GeV, and $m_b = 4.7$~GeV. $M'$ is fixed 
by the gaugino unification relation $M' = 
{\displaystyle{\frac{5}{3}}} \, \tan^2\tw M$ and the gluino mass is 
related to $M$ by $m_{\tilde g} = (\a_s(m_{\tilde g})/\a)\sin^2\tw 
M$. 
\\ 

\noindent {\bf Decays into stops:}

\begin{figure}[h!]
\begin{picture}(160,60)(0,0)
    %\graphpaper[5](0,0)(160,60)
    \put(30,7){\mbox{\resizebox{9cm}{!}
    {\includegraphics{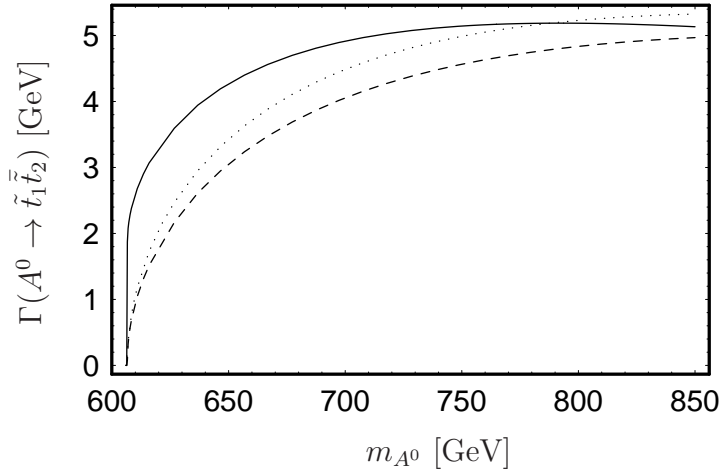}}}}
    \put(80,5){\makebox(0,0)[t]{{$m_{A^0}$ [GeV]}}}
    \put(23,20){\rotatebox{90}{{{$\Gamma (A^0 \rightarrow 
                                  \st_1 \!\bar{\,\st_2})$ [GeV]}}}} 
\end{picture}
\caption{Tree--level (dotted line), full electroweak corrected 
(dashed line) and full one--loop (electroweak and SUSY--QCD) 
corrected (solid line) decay width of $A^0 \rightarrow \st_1 
\!\bar{\,\st_2}$. \label{mA_dependence_os}} 
\end{figure}

In Fig.~\ref{mA_dependence_os} we show the tree--level and the 
corrected width to $A^0 \rightarrow \st_1 \!\bar{\,\st_2}$ for 
$\tan\b = 7$ and $\{ M_{\ti Q}, A, M, \mu \} = \{300, -500, 120, 
-260\}~\mbox{GeV}$ as a function of the mass of the decaying Higgs 
boson, $m_{A^0}$. As can be seen for larger values of $m_{A^0}$, 
the electroweak corrections can be of the same size as the 
SUSY--QCD corrections. 

%%%%%%%%%%%%%%%%%%%%%%%%%%%%%%% Fig 4 %%%%%%%%%%%%%%%%%%%%%%%%%%%%%%%%%
%%%%%%%%%%%%%%%%%%%%%%%%%%%%%%%%%%%%%%%%%%%%%%%%%%%%%%%%%%%%%%%%%%%%%%%

%%%%%%%%%%%%%%%%%%%%%%%%%%%%%%% Fig 5 %%%%%%%%%%%%%%%%%%%%%%%%%%%%%%%%%
\begin{figure}[h!]
\begin{picture}(160,65)(0,0)
    %\graphpaper[5](0,0)(160,65)
    \put(27,5){\mbox{\resizebox{9cm}{!}
    {\includegraphics{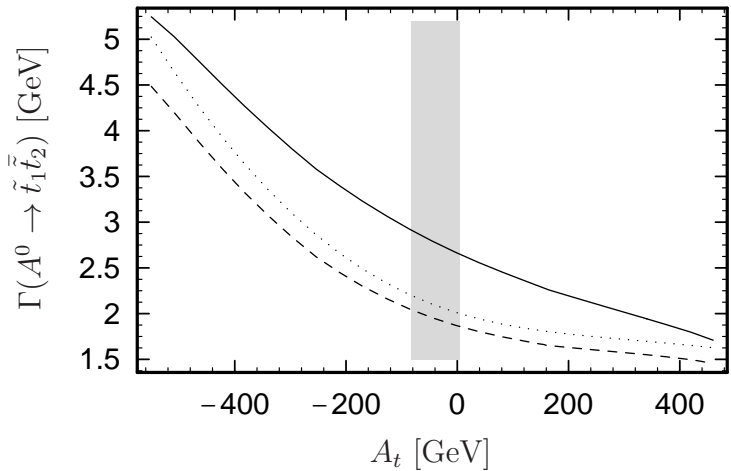}}}}
    \put(76,3){\makebox(0,0)[t]{{$A_t$ [GeV]}}}
    \put(20,20){\rotatebox{90}{{{$\Gamma (A^0 \rightarrow 
                                  \st_1 \!\bar{\,\st_2})$ [GeV]}}}} 
\end{picture}\caption{$A_t$--dependence of tree--level (dotted
line), full electroweak corrected (dashed line) and full one--loop 
(electroweak and SUSY--QCD) corrected (solid line) decay width of 
$A^0 \rightarrow \st_1 \!\bar{\,\st_2}$. The gray area is excluded 
by experimental bounds. \label{A_dependence_os}} 
\end{figure}
%%%%%%%%%%%%%%%%%%%%%%%%%%%%%%%%%%%%%%%%%%%%%%%%%%%%%%%%%%%%%%%%%%%%%%%

In Fig.~\ref{A_dependence_os} the tree--level, the full 
electroweak and the full one--loop corrected (electroweak and 
SUSY--QCD) decay width of $A^0 \rightarrow \st_1 \!\bar{\,\st_2}$ 
are given as a function of $A_t$. The electroweak corrections do 
not strongly depend on the parameter $A_t$ and are almost constant 
about 8\%. As input parameters we have chosen the values given 
above as well as $\{A_{b, \tau}, m_{A^0}\} = \{-500, 700\}$~GeV. 

\begin{figure}[h!]
\begin{picture}(160,65)(0,0)
    %\graphpaper[5](0,0)(160,65)
    \put(27,5){\mbox{\resizebox{9cm}{!}
    {\includegraphics{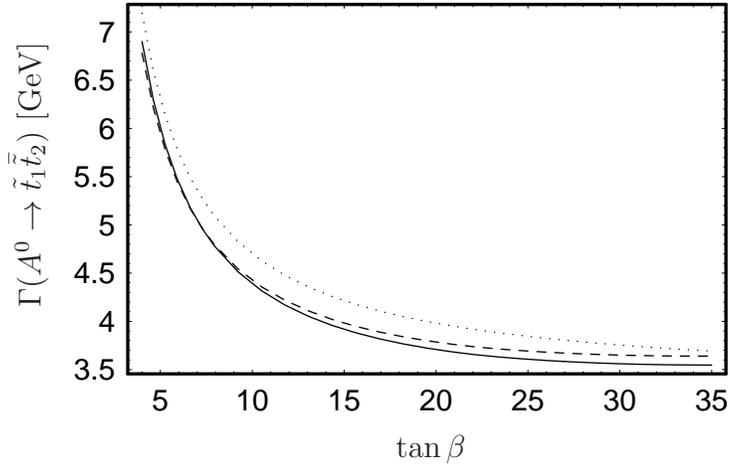}}}}
    \put(76,3){\makebox(0,0)[t]{{$\tan\b$}}}
    \put(20,20){\rotatebox{90}{{{$\Gamma (A^0 \rightarrow 
                                  \st_1 \!\bar{\,\st_2})$ [GeV]}}}} 
\end{picture}\caption{Tree--level (dotted line), full electroweak
corrected decay width (dashed line) and full one--loop 
(electroweak and SUSY--QCD) corrected width (solid line) of $A^0 
\rightarrow \st_1 \!\bar{\,\st_2}$ as a function of $\tan\b$. 
\label{tanbdep_st_os}} 
\end{figure}

Fig.~\ref{tanbdep_st_os} shows the tree--level, the full 
electroweak and the full one--loop corrected (electroweak and 
SUSY--QCD) decay width of $A^0 \rightarrow \st_1 \!\bar{\,\st_2}$ 
as a function of $\tan\b$ with the same parameter set as above and 
$m_{A^0}=900$~GeV. Again, in a large region of the parameter space 
the electroweak corrections are comparable to the SUSY--QCD ones. 
\\ 

%%%%%%%%%%%%%%%%%%%%%%%%%%%%%%% Fig 6 %%%%%%%%%%%%%%%%%%%%%%%%%%%%%%%%%
%%%%%%%%%%%%%%%%%%%%%%%%%%%%%%%%%%%%%%%%%%%%%%%%%%%%%%%%%%%%%%%%%%%%%%%

%%%%%%%%%%%%%%%%%%%%%%%%%%%%%%% Fig 7 %%%%%%%%%%%%%%%%%%%%%%%%%%%%%%%%%
\begin{figure}[h!]
\begin{picture}(160,65)(0,0)
    %\graphpaper[5](0,0)(160,65)
    \put(25,5){\mbox{\resizebox{9.45cm}{!}
    {\includegraphics{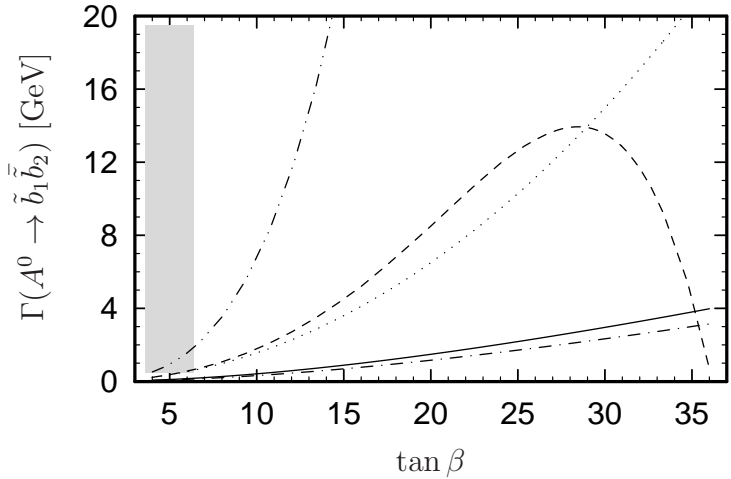}}}}
    \put(76,3){\makebox(0,0)[t]{{$\tan\b$}}}
    \put(20,20){\rotatebox{90}{{{$\Gamma (A^0 \rightarrow 
                                  \sb_1 \!\bar{\,\sb_2})$ [GeV]}}}} 
\end{picture}
\caption{Two kinds of perturbation expansion: the dotted line 
corresponds to the on--shell tree--level width, the dashed and 
dash--dot--dotted line correspond to electroweak SUSY--QCD 
on--shell one--loop width, respectively. The dash--dotted line 
corresponds to the improved tree--level and the solid line to the
(full) improved one--loop width.\label{tanb_sb_imp}} 
\end{figure}
%%%%%%%%%%%%%%%%%%%%%%%%%%%%%%%%%%%%%%%%%%%%%%%%%%%%%%%%%%%%%%%%%%%%%%%

\noindent {\bf Decays into sbottoms:}\\ Here we illustrate the 
numerical improvement of the full one--loop corrections to 
\mbox{$A^0 \rightarrow \sb_1 \!\bar{\,\sb_2}$} for large $\tan\b$. 

In Fig.~\ref{tanb_sb_imp} we show two kinds of perturbation 
expansion for the input parameters $\{m_{A^0}, M_{\ti Q}, A_t, 
A_b, A_\tau, M, \mu \} = \{800, 300, 150, -700, {-500}, 120,$ 
$260\}~\mbox{GeV}$: First we show the on--shell tree--level width 
(dotted line). The dashed and dash--dot--dotted line correspond to 
the on--shell electroweak and full (electroweak plus SUSY--QCD) 
one--loop width, respectively. For both corrections one can 
clearly see the invalidity of the on--shell perturbation 
expansion, in particular the electroweak corrections lead to an 
improper negative decay width. The second way of perturbation 
expansion is given by the dash--dotted and the solid line which 
correspond to the improved tree--level and improved full one--loop 
decay width, respectively. The smallness of the relative 
correction in this case shows that the improved tree--level is 
already a good approximation for \mbox{$A^0 \rightarrow \sb_1 
\!\bar{\,\sb_2}$}. The input parameters are the same as in the 
first case but now with running $A_b = -700~\mbox{GeV}$.

\noindent {\bf Squarks decays:}

\begin{figure}[h!]
\begin{picture}(160,65)(0,0)
    %\graphpaper[5](0,0)(160,65)
    \put(28.5,5){\mbox{\resizebox{9cm}{!}
    {\includegraphics{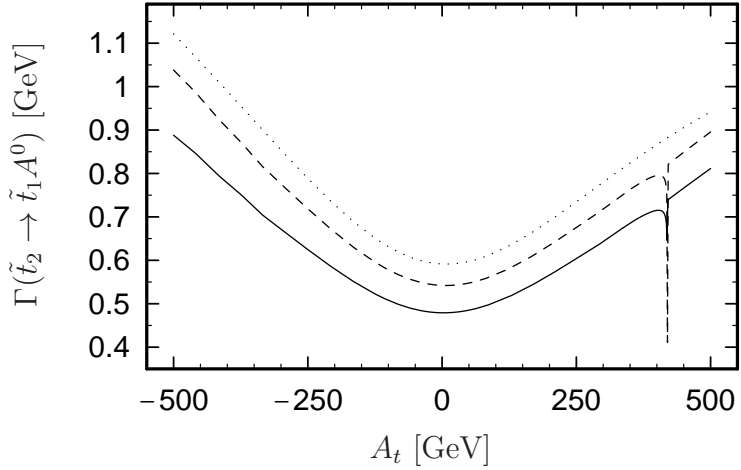}}}}
    \put(76,3){\makebox(0,0)[t]{{$A_t$ [GeV]}}}
    \put(20,20){\rotatebox{90}{{{$\Gamma (\st_2 \rightarrow 
                                  \st_1 A^0)$ [GeV]}}}} 
\end{picture}
\caption{$A_t$--dependence of the tree--level (dotted line), full 
electroweak corrected (dashed line) and full one--loop corrected 
(solid line) decay widths of $\st_2 \rightarrow \st_1 A^0$. 
\label{Atdep_st_os_crossed}} 
\end{figure}

Fig.~\ref{Atdep_st_os_crossed} displays the decay widths of the 
crossed channel $\st_2 \rightarrow \st_1 A^0$ as a function of 
$A_t$. As can be seen, the electroweak corrections are as large as  
the SUSY--QCD ones in the considered region. The 
values of the input parameters are $\{ \tan\!\b, \mu \} = \{ 35, 
-300 \}$ and $\{m_{A^0}, m_{\ti g}, M_{\ti Q}, A_b, A_\tau \} = 
\{150, 1000, 300, -700, -700 \}~\mbox{GeV}$ with the relations for 
the SUSY breaking masses given at the top of this section but with 
$M_{{\ti U}_3} = 500~\mbox{GeV}$ in order to get a quite 
acceptable mass splitting in the stop sector. 

%%%%%%%%%%%%%%%%%%%%%%%%%%%%%%% Fig 8 %%%%%%%%%%%%%%%%%%%%%%%%%%%%%%%%%
%%%%%%%%%%%%%%%%%%%%%%%%%%%%%%%%%%%%%%%%%%%%%%%%%%%%%%%%%%%%%%%%%%%%%%%

%%%%%%%%%%%%%%%%%%%%%%%%%%%%%%% Fig 9 %%%%%%%%%%%%%%%%%%%%%%%%%%%%%%%%%
\begin{figure}[h!]
\begin{picture}(160,65)(0,0)
    %\graphpaper[5](0,0)(160,65)
    \put(28.5,5){\mbox{\resizebox{9cm}{!}
    {\includegraphics{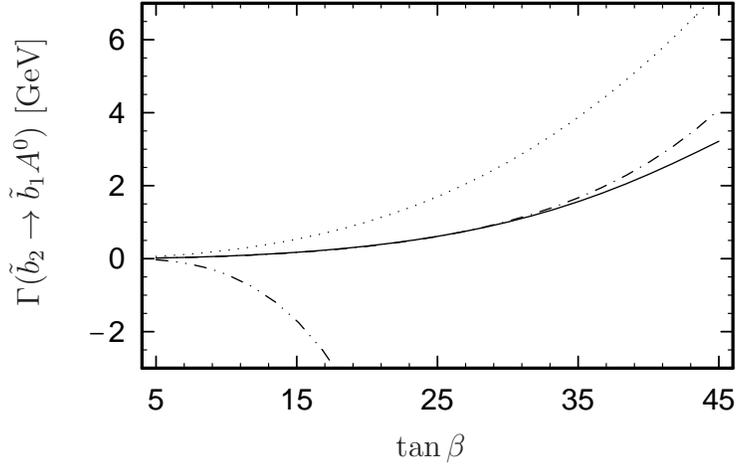}}}}
    \put(76,3){\makebox(0,0)[t]{{$\tan\b$}}}
    \put(20,20){\rotatebox{90}{{{$\Gamma (\sb_2 \rightarrow 
                                  \sb_1 A^0)$ [GeV]}}}} 
\end{picture}
\caption{Decay widths of $\sb_2 \rightarrow \sb_1 A^0$ as a 
function of $\tan\b$. The dotted and dash--dot--dotted line 
correspond to the on--shell tree--level and on--shell one--loop 
width, respectively. The dash--dotted line corresponds to the full 
improved tree--level and the solid line to the full improved 
one--loop width. \label{tanb_sb_imp_crossed}} 
\end{figure}
%%%%%%%%%%%%%%%%%%%%%%%%%%%%%%%%%%%%%%%%%%%%%%%%%%%%%%%%%%%%%%%%%%%%%%%

Fig.~\ref{tanb_sb_imp_crossed} again demonstrates the numerical 
improvement in the large $\tan\!\b$ regime: The dotted and 
dash--dot--dotted lines correspond to the on--shell tree--level 
and on--shell one--loop decay widths of $\sb_2 \rightarrow \sb_1 
A^0$, whereas the dash--dotted and solid lines show the full 
improved tree--level and one--loop widths, respectively. The input 
parameters are the same as in Fig.~\ref{Atdep_st_os_crossed} but 
with $\{ M_{{\ti Q}_3}, A \} = \{ 500, -700 \}~\mbox{GeV}$.

\vspace{2mm} In conclusion, we have calculated the {\em full} 
electroweak one--loop corrections to the decay widths $A^0 
\rightarrow \tilde{q}_1 \ {\bar{\!\tilde{q}}}_{2}$ and 
$\tilde{q}_2 \rightarrow \tilde{q}_1 A^0$ in the on--shell scheme. 
Moreover, we have included the SUSY--QCD corrections which were 
calculated in \cite{SUSY-QCD}. For the decay into sbottom quarks 
and large $\tan\b$ an improvement of the on--shell perturbation 
expansion is necessary. This was done by an appropriate 
redefinition of the tree--level Higgs--squark--squark coupling. We 
find that the corrections are significant and in a wide range of 
the parameter space comparable to the SUSY--QCD corrections.\\

\noindent {\bf Acknowledgements}\\ \noindent The authors 
acknowledge support from EU under the HPRN-CT-2000-00149 network 
programme and the ``Fonds zur F\"orderung der wissenschaftlichen 
Forschung'' of Austria, project No. P13139-PHY.

\end{document}